\documentstyle[prl,aps,epsf]{revtex}
\begin{document}
\advance\textheight by 0.5in
\advance\topmargin by -0.25in
\draft

\twocolumn[\hsize\textwidth\columnwidth\hsize\csname@twocolumnfalse%
\endcsname

\title{Broken Symmetries in  Scanning Tunneling Images of Carbon Nanotubes}
\author{C.L. Kane and E.J. Mele}
\address{Department of Physics \\
Laboratory for Research on the Structure of Matter \\
 University of Pennsylvania\\
Philadelphia, Pennsylvania 19104 }

\date{\today}
\maketitle

\begin{abstract}
Scanning tunneling images of carbon nanotubes frequently show electron
 distributions which break the local sixfold symmetry of the graphene sheet.
 We present a theory of these images  which relates these anisotropies to the off
 diagonal correlations in the single particle density matrix, and  allows one to
 extract these correlations from the observed images. The theory is 
 applied to images of the low energy states reflected at the end of a tube or
 by point defects, and to states
 propagating on defect free semiconducting tubes.  The latter exhibit
 a novel switching of the anisotropy in  the tunneling image with the sign
 of the tunneling bias.  
\end{abstract}
\vskip -0.5 truein
]

Scanning tunneling microscopy and spectroscopy is a powerful
 tool for studying the structural and electronic properties of carbon 
 nanotubes at the atomic scale. Several experimental groups have  reported
 tunneling images of isolated single wall carbon nanotubes\cite{dekker1,lieber}   and of
 tubes packed into bundles or ``ropes"\cite{cj}. In some cases these measurements
 have allowed a direct determination of the diameters and wrapping vectors
 for the tubes and these observations, combined with scanning tunneling spectroscopy,
  have confirmed the idea that the semiconducting
 or conducting behavior of a tube is controlled by its wrapping vector\cite{dekker1,lieber}. 

 However STM images of these systems obtained at 
low bias voltages contain a number 
 of surprising features. At low bias voltages the images rarely 
display the full sixfold lattice symmetry even when the underlying
graphene lattice is undistorted.
 Instead,  these images frequently contain
 a broken symmetry in the form of ``striped"  patterns in 
which maxima in the electron density are observed in bond chains which spiral around
 the tube\cite{dekker1,lieber,cj}. 
 In some images superlattice structures are present
with a period commensurate with but larger
 than that of the underlying graphene sheet\cite{cj2}.  
Moreover, energy resolved images of short tubes show standing waves 
characteristic of individual eigenstates which also have
a period longer than that of the graphene lattice\cite{dekker2}.


 In special cases broken translational or rotational symmetries 
obtained in an STM image can be attributed to
 asymmetries in the tunneling tip\cite{artifact}.   In this Letter we point out that
 asymmetric images are expected  even in an \it ideal \rm tunneling experiment
 and contain important information about the low lying electronic states in
 these systems.  The asymmetries are \it interference patterns \rm
which are sensitive to the  coherence between the ``forward" and ``backward"
 moving electronic states propagating on the tube walls. This can arise from backscattering from
 tube ends, from various defects on the tube walls, and even from propagation in a translationally
 invariant potential on a semiconducting tube. We show that these
 interference patterns in the tunneling images are fingerprints which
directly probe the  off diagonal correlations in the
 single particle density matrix.  We provide a theory for extracting these correlations
 from the observed images.  This effect is illustrated with several examples 
 of the tunneling densities of states calculated for 
 defect free tubes and for tubes with point defects. 
  
 The low lying electronic states on a carbon nanotube are derived from the 
 propagating states near the ``Fermi points" of an isolated graphene sheet
 located at the Brilloun zone corners shown in the lower panel of Fig. 1\cite{estructure,km}.
 There are two inequivalent points which we will refer to as
 K = $K_0$ and K'=-$K_0$. 
Wrapping the graphene sheet into a cylinder requires that the electronic
wavefunctions satisfy periodic boundary conditions
$\Psi(\vec{r} + \vec{T}_{m,n}) = \Psi(\vec{r})$ where
$\vec{T}_{m,n} = m \vec{\tau}_1 + n\vec{\tau}_2$  gives
the wrapping vector around the tube circumference  expressed  in terms of
the two primitive graphite translation vectors
 $\vec{\tau}_1 = (1/2,\sqrt{3}/2)$ and $\vec{\tau}_2 = (-1/2,\sqrt{3}/2)$.
When $ {\rm mod} (m-n,3) = 0 $ the Bloch waves \it exactly at \rm
the K and K' points are allowed quantized waves on the tube.  This leads
to the metallic band structure in Fig. 1(a).  In contrast, when
${\rm mod}(m-n,3) = (1,2) $ the allowed quantized waves do not
intersect K and K', which leads to a semiconducting gap in the electronic
spectrum as shown in Fig. 1(b).

 The spectrum of Fig. 1(a) describes the propagating modes of a defect free conducting
 tube. However, these waves can be reflected from tube ends or from defects along the tube.
 The interference between the forward and backward moving  waves produces  a spatial
 modulation of the charge density. Since the carbon nanotubes have two forward and backward
 moving channels (associated with the K and K' points shown in Fig. 1 the 
resulting interference patterns
 have a particularly rich structure. 
 The coherent superposition 
 of the forward and backward moving components of the scattering
 states produces off diagonal correlations in the density
 matrix at energy $E$,
\begin{equation}
\rho_{\alpha\beta}(E) =
\langle \psi^\dagger_\alpha \delta( E - {\cal H})
\psi_\beta \rangle.
\end{equation}
where ${\cal H}$ is the Hamiltonian and $\alpha$ is a four component index 
specifying the left and right moving bands at the K and K' points.

 To derive the local tunneling density of states from the  density matrix in 
equation (1) we represent the  
 Bloch waves $\psi_\alpha(\vec{r})$ as a sum atomic orbitals
 centered on sites $\vec{\tau}_m$ in cells $\vec{T}_n$
\begin{equation}
\psi_{\alpha}(\vec{r}) = \sum_{m,n} \gamma_{m \alpha}
 e^{i \vec{k}_{\alpha} \cdot \vec{T_n}} f(\vec{r} -
\vec{\tau}_m - \vec{T}_n).
\end{equation}
where $\gamma_{m \alpha}$ are the amplitudes for the Bloch state on sites $m$.
These Bloch waves can be represented as an expansion in reciprocal
 lattice vectors
\begin{equation}
\psi_{\alpha} = \sum_{m,n} \gamma_{m \alpha} e^{-i(\vec{k}_{\alpha} 
+ \vec{G}_n) \cdot \vec{\tau}_m} F(|\vec{k}_{\alpha} +
\vec{G}_n |) e^{i (\vec{k}_{\alpha} + \vec{G}_n) \cdot \vec{r}}.
\end{equation}
For tunneling from a tip which is smooth on a scale of the
atomic spacing, $F(q)$ decreases rapidly for large $q$.
In the following we will truncate this expansion, keeping only $\vec G$'s
in the lowest ``star" of $\vec k_\alpha + \vec G$.  
This becomes exact when the STM tip is sufficiently  
high above the surface\cite{tersoff}.
Including higher Fourier components does not significantly change our conclusions.
We also assume the tip is isotropic, so within the first 
star $F(|\vec k + \vec G|)$ is independent of $\vec G$.

The local density of states at energy $E$ can be expressed
\begin{equation}
\rho(\vec r , {\rm E}) = \psi^* _{\alpha} (\vec r) \rho_{\alpha \beta}(E) \psi _{\beta} (\vec r).
\end{equation}
It is useful to characterize the tunneling image in terms of its 
longest wavelength Fourier components.  Coupling between bands
at the same K point leads to images with the periodicity of
the lattice.  These are described by Fourier components in the
first star of reciprocal lattice vectors $\vec q_{1i} = \vec G_i$, indicated in 
Fig. 1.  On the other hand coupling between the two K points
leads to modulations with a $\sqrt{3} \times \sqrt{3}$ superlattice,
which are described by the ``$\sqrt{3}$" star, $\vec q_{\sqrt{3}i} = \vec K_i$.
The discussion is further simplified by projecting onto the ``triangular harmonics"
defined in each star to be
\begin{equation}
\rho_{p m}  = {1\over 3}\sum_{n=0}^2 e^{2\pi i m n/3}
\int d^2 r e^{-i \vec q_{p n} \cdot \vec r}\rho(\vec r)
\end{equation}
with $p = 1$ or $\sqrt{3}$ and $m = -1,0,1$.
Combining (1-5) yields a simple expression for the 
Fourier components $\rho_{p m}(E)$ in
terms of the density matrix $\rho_{\alpha\beta}(E)$.

We first consider the effects of reflection either from the end of 
the tube or from
an impurity.  The reflection of waves associated with the K and K' points is
characterized by a $2\times 2$ matrix of complex reflection amplitudes. 
This matrix contains three independent amplitudes labeled $r_a$, $r_b$ and 
$r_m$ in Fig. 1(a). These describe respectively large momentum scattering
 from $K \rightarrow K'$, $K' \rightarrow K$ and small momentum scattering at the $K$ and $K'$ points.
The equality between the two amplitudes described by $r_m$ follows from time
reversal symmetry.  These reflection amplitudes depend on the detailed structure of the 
scatterer, although for special high symmetry scatterers  its form can be constrained by symmetry. 
In the following, we consider an infinitely long tube, so that the density matrix
is diagonal in the basis of scattering states, which are a superposition of incoming and
reflected waves.

The large momentum scattering amplitudes $r_a$ and $r_b$ illustrated in
Fig. 1 produce a modulation of the TDOS in the $\sqrt{3}$ star.
For a conducting tube we find that the TDOS measured at energy E and at 
a distance x from a point scatterer has the Fourier  coefficients\cite{origin}
\begin{equation}
\begin{array}{rcl}
\rho_{\sqrt{3}0}(E) &=& {1\over 2}N(E)(r_b e^{2iQx} - r_a^* e^{-2iQx})
\\ \\
\rho_{\sqrt{3}\pm 1}(E) &=&
{1\over 2} N(E)e^{\pm i \theta} (r_b e^{2iQx} + r_a^* e^{-2iQx}) \\
\end{array}
\end{equation}
where $N(E)$ is the density of states, $Q=E/ \hbar v_F$, $v_F$ is
 the Fermi velocity and $\theta$ is the chiral angle which orients
 the zigzag bond direction with respect to the tube axis (i.e.
 $\theta = 0$ defines a tube with an armchair wrapping).  The
 dependence of the TDOS on $r_a$ and $r_b$ is most clearly
 seen for tunneling at low bias ($E \approx 0$). The $Q$ dependence in equation (5) 
arises from the fact that the scattered states
at nonzero energy are not  located precisely at the Fermi points.

As an example in Fig. 2 we display the calculated TDOS at E=0 for
the values $r_a=r_b=1$ on an armchair tube. The scattering amplitudes produce a 
$\sqrt{3} \times \sqrt{3}$ modulation of the  tunneling image which 
in which the bond charges are enhanced in a superlattice of bonds 
oriented along the circumferential direction of the tube. 
Similar $\sqrt{3} \times \sqrt{3}$ modulations occur in the presence
of impurities on the surface of graphite\cite{mizes2}.   Those patterns 
follow from the two dimensional scattering between the $K$ and $K'$
points in the graphite plane.

 The small momentum backscattering amplitudes $r_m$  produce
 \it cell periodic \rm modulations of the TDOS which can nonetheless break
 the rotational symmetry of the image.  These effects are   
 produced by a modulation of the  Fourier components of the TDOS in the first
 star of reciprocal lattice vectors $\vec G_n$ shown in Fig. 1. Using the
 expansion in triangular harmonics in equation (5) we find\cite{origin}
\begin{equation}
\begin{array}{rcl}
\rho_{10}(E) &=& N(E)(-1 + i\sqrt{3} {\rm Re}[r_m e^{2iQx}]) \\ \\
\rho_{1\pm 1}(E) &=& \mp i N(E) e^{\pm i\theta}
{\rm Im}[r_m e^{2iQx}]. \\
\end{array}
\end{equation}
 Interesting structure in the TDOS is produced by the imaginary part of $\rho_{1 \pm 1}$. 
This generally occurs for any chiral tube
 with $\theta \ne 0$, but it can \it also \rm  occur for a nonchiral armchair
 tube, $\theta = 0$, when the $q \approx 0$ backscattering amplitude $r_m$
 develops a nonvanishing imaginary part. This leads to the interference pattern
 shown in Fig. 2(b), in which a bond density
 wave is deflected into a spiral pattern around the 
 axis of the armchair tube.  We  find that a  reflection amplitude with this symmetry can be
 produced by any point defect or end cap which breaks the two sublattice
 symmetry of the  underlying graphene sheet. 

Semiconducting tubes with ${\rm mod}(m-n,3) \ne 0$ have a gap in the low energy spectrum as shown
 in the middle panel of Fig. 1. This gap arises from a  coherent superposition of forward and backward 
 moving components which is required for the wavefunction to satisfy 
periodic boundary conditions  around the tube waist. 
 Exactly at the band edges one obtains a perfect standing wave which contains an equal admixture
 of forward and backward propagating waves. 
 Note however that the backscattering responsible for  these states does not result from  reflection
 from an isolated point defect, but
 instead arises from the presence of a ``mass" operator in the low energy Hamiltonian\cite{km}
 which  preserves the
 lattice translational symmetry, and  \it breaks  \rm its rotational symmetry. Remarkably, 
the symmetry of this ``mass"
 term depends sensitively on the wrapping vector and allows one to distinguish ${\rm mod}(m-n,3)=1$ from
 ${\rm mod}(m-n,3)=2$ tubes.  To do this we define the chiral index $s = (-1)^{{\rm mod}(m-n,3)}$. 
Then we find that the tunneling density
 of states for a semiconducting tube with chiral index $s$, chiral angle $\theta$ and gap
 2$\Delta$ is given by\cite{origin}
\begin{equation}
\begin{array}{rcl}
\rho_{10}(E)  &=& - N(E) \\ \\
\rho_{1\pm 1}(E) &=& \pm i s N(E) e^{\pm i\theta} \Delta/E. \\
\end{array}
\end{equation}
 In Fig. 3 we display tunneling densities of states calculated for  the band edge states 
$(E = \pm \Delta)$ on tubes with
  wrapping indices [m,n] =  [12,7], [14,6] and [17,0] (zigzag).  These three tubes have a common chiral
 index $s$=1 and chiral angles which vary between $8.6^\circ $ and $30^\circ$. The band edge states imaged at
 positive and negative bias have \it complementary  \rm structures, that is the  superposition of the two images gives an
 image with perfect sixfold symmetry although each image separately breaks the sixfold symmetry. 
 For chiral angles near $\theta \approx 0$ the tunneling images consist of a series of complementary 
 spiral stripes.  However the TDOS changes smoothly as a function of chiral angle, so that as  
 one approaches the zigzag structure the negative energy states are enhanced in  a pattern of 
 isolated bonds in the structure, while the density for the positive energy 
states is confined to a connected zigzag
 bond chain. Note that the symmetry breaking terms in $\rho_{1 \pm 1}$ depend on the product of the chiral
 index \it and \rm the energy. Thus for a [12,8] tube which has a  chiral index -1 the symmetries of the
 positive and negative energy solutions are reversed. 

 Fig. 3 shows that for semiconducting tubes the tunneling images obtained near the band
 edges break the local point symmetry of the graphene sheet, 
and that for a given tube the  sign of the symmetry
 breaking is switched by reversing the bias of the tunneling tip.  
Observation of such a reversal
 in the tunneling image would provide a striking identification of the chiral index $s$ of a
 semiconducting tube, even when the wrapping indices [n,m] can not be resolved.
Moreover this reversal would clearly distinguish this effect from a tip artifact\cite{artifact}.
 We should note
 that several experiments have already suggested that a tube in contact with a conducting substrate can
 be doped ``p" type \cite{dekker1} and so  
one needs to correct such a measurement for the offset in the chemical 
 potential for any such  unintentionally  ``doped" tube. 
One also needs to ensure that the tunneling is carried out in 
 an energy window where only one azimuthal subband is accessible, since the symmetries of
 the tube eigenstates alternate in successive conduction or valence subbands, 
 tending to suppress the anisotropy in the tunneling images.

 The data presented in Figures 2 and 3 demonstrate that scanning tunneling images of carbon nanotubes
 are very sensitive to the nodal structure of the underlying electronic states\cite{tersoff}. 
 Indeed at low energy these images are probing the internal structure of \it individual \rm 
(or at best a small number) of electronic eigenstates on the tube surface. The 
 appearance of these broken symmetry patterns in a tunneling image 
does not imply large structural perturbations to the
 covalent graphene network. Indeed the data presented here are calculated for structures which are \it
 all \rm  unstrained and  perfectly locally sixfold symmetric.  These density patterns do 
require coherence between forward and backward propagating waves.  In fact  a  quantitative
 analysis of these images can be used to extract off diagonal correlations in the density matrix
 in equation (1)  responsible for these patterns.  These data can then be used to extract the
scattering matrix, which characterizes the internal
 structure and symmetry of various scattering centers, and to identify the wrapping vector 
 for semiconducting tubes.   Finally, we remark that the off diagonal correlations 
responsible for these patterns could also arise due to 
interactions with a substrate, interactions between tubes in a 
rope or due to electron-electron interactions.


It is a pleasure to thank W. Clauss and A.T. Johnson for helpful discussions of their
STM images.  This work has been supported by the NSF under grants
DMR 95-05425, DMR 96-32598 and DMR 98-02560 
and by the DOE under grant DE-FG02-84ER45118.

\begin{figure}
   \epsfxsize=1.8in
\centerline{\epsfbox[141 388 267 690]{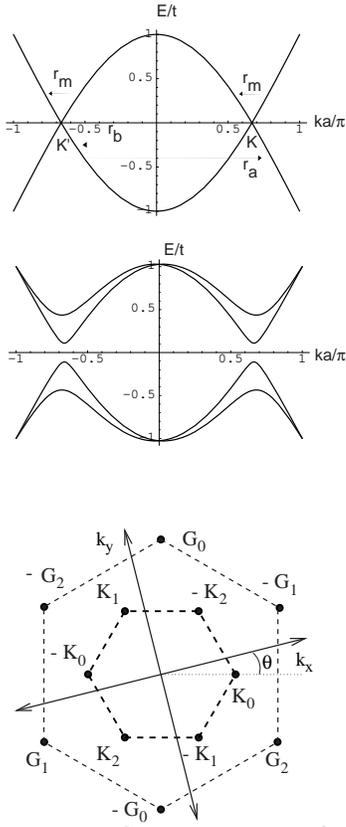}}
   \caption{Low energy electronic spectra for a conducting tube (top)
and for a semiconducting tube (middle). The lower panel shows the momenta
 in the first two stars in reciprocal space which describe the local tunneling
 density of states in an STM experiment.}
\end{figure}

\begin{figure}
   \epsfxsize=3in
   \centerline{\epsffile{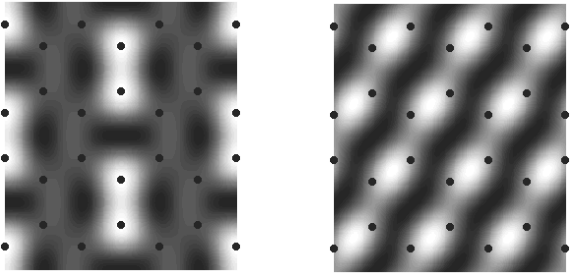}}
   \caption{Local tunneling densities of states for a [10,10] tube
 in the presence of reflection from point defects. The left panel shows a
 tunneling image which is modulated in a $\sqrt{3} \times \sqrt{3}$
 pattern and in the right panel shows a ``primitive" $1 \times 1$ spiral striped pattern which breaks a mirror symmetry of the armchair tube.}
\end{figure}

\begin{figure}
   \epsfxsize=3in
   \centerline{\epsffile{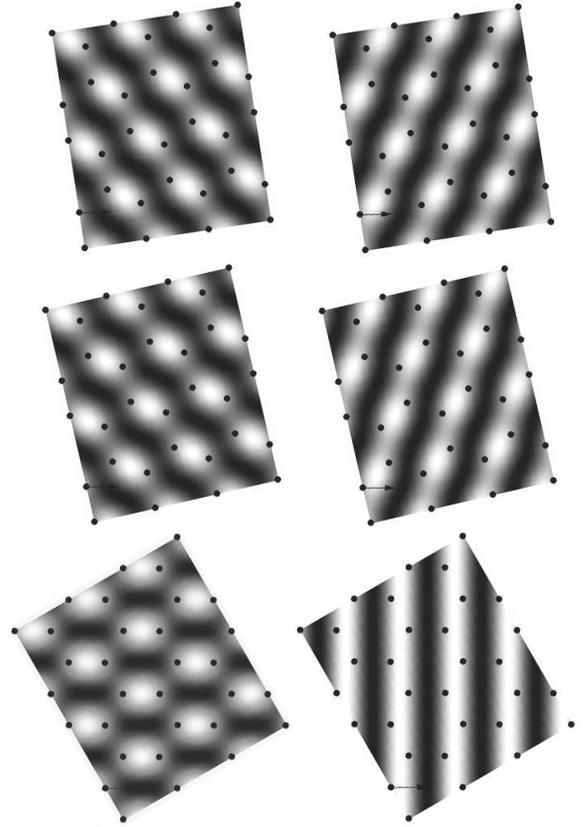}}
   \caption{Complementary tunneling densities of states for the two band edges
 of three semiconducting chiral tubes with the same chiral index. The left panel gives the image
 for tunneling into a band edge state at negative energy $(E = -\Delta)$, and the right panel
 gives the corresponding image at positive energy $(E= \Delta)$. The images are for tubes with
 wrapping indices [12,7] (top), [14,6] (middle) and [17,0] (bottom). In each
 case the image is rotated so that the tube axis is directed along the horizontal.}
\end{figure}


\begin{references}
\bibitem{origin}
To obtain these formulas we specify that the origin ${\vec r} = 0$
lies a the center of a hexagon.
\end{references}
\end{document}